\documentclass[a4paper,11pt]{article}
\usepackage{jcappub}

\usepackage{siunitx}
\usepackage{graphicx}
\usepackage{bm}
\pdfoutput=1 
\usepackage{color}
\usepackage{makecell}
\usepackage{afterpage}
\usepackage[dvipsnames]{xcolor}
\usepackage{multirow}
\usepackage{cleveref}
\usepackage{subfigure}
\usepackage{booktabs}
\usepackage[normalem]{ulem}		
\usepackage{pgfgantt}
\usepackage{ulem}

\newcommand{\be}{\begin{equation}}
\newcommand{\ee}{\end{equation}}

\newcommand{\Nm}{{\rm Nm}}
\newcommand{\N}{{\rm N}}

\newcommand{\CLASS}{{\sc class}}
\newcommand{\gevolution}{\textit{gevolution}}
\newcommand{\GADGET}{{\sc gadget-4}}

\newcommand{\HL}{H_{\rm L}}
\newcommand{\HT}{H_{\rm T}}
\newcommand{\HTp}{\dot{H}_{\rm T}}
\newcommand{\HTpp}{\ddot{H}_{\rm T}}

\title{A Minimal Model for Massive Neutrinos in Newtonian N-body Simulations}

\author[1]{Pol Heuschling,}

\emailAdd{pol.heuschling@rwth-aachen.de}

\author[2]{Christian Partmann,}

\emailAdd{partmann@mpa-garching.mpg.de}

\author[1]{Christian Fidler}

\emailAdd{fidler@physik.rwth-aachen.de}

\affiliation[1]{Institute for Theoretical Particle Physics and Cosmology (TTK), \\ RWTH Aachen University, D-52056 Aachen, Germany.}

\affiliation[2]{Max-Planck-Institut für Astrophysik (MPA), Karl-Schwarzschild-Str. 1, D-85748, Garching, Germany.}

\abstract{
We present a novel method for including the impact of massive neutrinos in cold dark matter N-body simulations. Our approach is compatible with widely employed Newtonian N-body codes and relies on only three simple modifications. First, we use commonly employed backscaling initial conditions, based on the cold dark matter plus baryon power spectrum instead of the total matter power spectrum. Second, the accurate Hubble rate is employed in both the backscaling and the evolution of particles in the N-body code. Finally, we shift the final particle positions in a post-processing step to account for the integrated effect of neutrinos on the particles in the simulation. However, we show that the first two modifications already capture most of the relevant neutrino physics for a large range of observationally interesting redshifts and scales. The output of the simulations are the cold dark matter and baryon distributions and can be analysed using standard methods. All modifications are simple to implement and do not generate any computational overhead. By implementing our methods in the N-body codes \GADGET{} and \gevolution{}, we show that any state-of-the-art Newtonian N-body code can be utilised out of the box. Our method is also compatible with higher order Lagrangian perturbation theory initial conditions and accurate for masses up to at least $\sum m_\nu = 0.3 \, \si{eV}$. Being formulated in relativistic gauge theory, in addition to including the impact of massive neutrinos, our method further includes relativistic corrections relevant on the large scales for free.}

\begin{document}

\maketitle

\section{Introduction}

Future generations of large-scale structure surveys, such as DES \cite{DES_2020}, DESI \cite{DESI_2016}, EUCLID \cite{euclid} and LSST \cite{lsst} will provide highly accurate measurements of the clustering of matter on large scales and in particular the matter power spectrum. To optimally exploit these measurements for high-precision test of the standard model of cosmology, it is necessary to improve the accuracy of numerical simulations to match the experimental sensitivity. This necessarily includes the modelling of small, sub-leading effects such as the dynamics of massive neutrinos and other possible extensions to the $\Lambda$CDM model.

In this paper, we focus on accurately including the impact of massive neutrinos in N-body simulations, which leave a measurable imprint on the clustering of matter on large scales, i.e. a scale-dependent suppression of the growth of structure. This mass dependent feature opens a window into investigating the nature of neutrinos, potentially revealing the absolute mass scale and type of hierarchy into witch the neutrinos are organised. 

Computing the rich dynamics of non-linear structure formation is a challenging task, even without massive neutrinos. While it is relatively efficient to employ N-body simulations for cold dark matter, including the relativistic and hot neutrinos often requires significant additional computational resources. In this paper we present a novel method for including the impact of massive neutrinos on the cold dark matter evolution that only minimally modifies existing cold dark matter simulations and therefore does not significantly increase the computational complexity. The method works with any N-body code and does not interfere with commonly employed numerical optimisations.    

\subsection{Neutrino physics}
\label{sec:neutrino_physis}
Advances in modern particle physics have enabled us to predict a non-zero neutrino mass due to the fact that neutrinos oscillate between their flavours. Experiments like T2K and Kamland have observed those oscillations \cite{de_Salas_2018} and found a lower bound of $ \sum_{\nu} m_\nu = M_{\nu} > 0.06 \si{.eV}$ at $95 \% $ confidence level. Another approach to probe the mass is the measurement of the energy spectrum of a beta-decay of e.g. tritium. One of such experiments, KATRIN, found an upper bound of $m(\nu_e)  < 1.1 \si{.eV}$ at $90 \% $ confidence, see \cite{Aker_2019}. From a completely different perspective, cosmology offers various ways to measure the neutrino mass. Instead of directly testing the properties of neutrinos in a laboratory, we can observe the effect of the tiny gravitational forces that neutrinos apply on the dark matter and baryons populating the Universe. While the force exerted by a single neutrino is small, in cosmology we can observe their collective interaction integrated over timescales as large as the age of the Universe, imprinting the structures in our Universe with some characteristic and mass-sensitive features. Using the high precision measurement of CMB anisotropies, the 2015 data release of PLANCK could constrain the mass to $M_{\nu} < 0.537 \si{.eV}$ at $95 \% $ level or even to $M_{\nu} < 0.257 \si{.eV}$ at $95 \% $ level, when polarisation data is included \cite{Planck_2015}. An even more stringent bound can be found analysing the 1D Lymann-$\alpha$ power spectrum \cite{Palanque_Delabrouille_2015}, or by joining CMB, BAO and high $l$ polarisation data leading to a bound of roughly $M_{\nu}<0.12 \si{.eV}$ at $95 \% $ C.L. \cite{vagnozzi2017unveiling}. A later data release of the PLANCK experiment \cite{Planck_2018} found the same bound of $M_{\nu}<0.12 \si{.eV}$ at $95\%$ C.L. using CMB and BAO data only. Future large-scale structure surveys are expected to significantly tighten those bounds eventually reaching the mass range required to explain the observed neutrino oscillations.

Neutrinos have a couple of important properties that set them apart from cold dark matter concerning structure formation, as explained in detail in e.g. \cite{Lesgourgues:2013sjj}. As neutrinos carry a non-zero mass, they contribute to the massive species of the Universe. Their present energy density in units of the critical density is given as
\begin{equation}
\Omega_{\nu} = \frac{\sum_{\nu} m_{\nu}}{93.14 \si{.eV.} h^2 } \,.
\end{equation}
Because of their very small mass, neutrinos travel at large velocities. At very early times, neutrinos are relativistic, but the expansion of the universe cools them down such that they transition into the non-relativistic regime around redshift
\begin{equation}
z_{\rm nr} = \frac{m_{\nu}}{5.28 \cdot 10^{-4} \si{.eV}} -1 .
\end{equation}

Before this transition, the spatially averaged neutrino density $\rho_{\nu}(a)$ scales similar to radiation $\rho_{\nu}(a) \propto a^{-4}$ while for $z <z_{\rm nr}$, the neutrino density evolution tends to $\rho_{\nu}(a) \propto a^{-3}$, as expected for a cold matter species. 
After their non-relativistic transition, the neutrino velocities are still orders of magnitude larger than e.g. CDM velocities, around
\begin{equation}
v_{\rm thermal}^{\nu} = 158(1+z) \left( \frac{\si{eV}}{m_{\nu}} \right) \si{km.s}^{-1}\,.
\end{equation}
For the allowed mass range of $M_{\nu}<0.12 \si{.eV}$ this gives a typical neutrino velocity today of about $1300 \si{.km .s^{-1}} $, which is roughly an order of magnitude faster than CDM particles moving at typical velocities of $\mathcal{O}(100 \si{.km.s^{-1}} )$, while the difference is much larger at earlier times. 
Consequently, neutrinos have a large free streaming scale
\begin{equation}
 \lambda_{\rm FS} = 8.10(1+z) \frac{H_0}{H} \left( \frac{\si{eV}}{m_{\nu}} \right) \si{Mpc.} h^{-1} .
\end{equation}
On large scales $\lambda \gg  \lambda_{\rm FS}$ neutrinos behave largely similar to CDM while perturbations on smaller scales are erased by free streaming. Even though neutrinos start to cluster on small scales in the late Universe, for realistic neutrino masses their growth is significantly delayed compared to ordinary matter and their dynamics remains well described by linear theory, even on small scales \cite{Adamek_2017}.

A convenient way to analyse the impact of massive neutrinos is their distinctive imprint on the matter power spectrum, where we consider a fixed value of $\omega_{\rm m}$ such that an increased neutrino mass leads to a reduction of $\omega_{\rm cdm}$. From the considerations above, one can estimate that three distinct effects will influence the matter power spectrum. First off, small scale neutrino perturbations are erased by free-streaming. However, more important is the delay of matter-radiation equality and a reduced growth rate of CDM at late times due to a missing clustering component \cite{Lesgourgues_2012}.

As a result, the power spectrum on small scales can be suppressed by several percent depending on the considered neutrino mass. From theoretical arguments, a suppression of the linear total matter power spectrum of $\Delta P / P \vert_{ \rm lin} \approx -8 f_{\nu}$ was found in \cite{Hu_1998}, where $f_{\nu}$ is the neutrino density fraction w.r.t. the total matter density. In non-linear simulations, this suppression is enhanced and $\Delta P / P \approx -9.8 f_{\nu}$ was found with a suite of N-body simulations \cite{Brandbyge_2008}. The additional non-linear suppression is expected, since the CDM growth is suppressed already in linear theory and, as a consequence, CDM perturbations also reach the non-linear regime a bit later \cite{Lesgourgues:2013sjj}. 

In the light of future experiments, an accurate modelling of neutrinos in dark matter simulations is essential, and a variety of different methods has been proposed in the literature. One technique is to include massive neutrinos directly, as particles, in N-body simulations \cite{Brandbyge_2008, Viel_2010, Adamek_2017, Villaescusa_Navarro_2013}. However, this approach is numerically challenging for various reasons. For instance, the high neutrino velocities require a description in special relativity \cite{Nascimento_2021}. Furthermore, since neutrinos do not cluster significantly and hence occupy a large volume in phase-space (in contrast to the dark matter sheet), it is difficult to sample the phase-space accurately without being limited by shot noise. Typically, the shot noise can only be reduced by an increase in the particle number, which in turn increases the computational cost significantly (however, see e.g. \cite{Banerjee2018, Elbers2021} for recent attempts to reduce the noise). On the other hand, this methods, if used with a sufficiently large neutrino particle number, naturally also compute non-linear corrections to the neutrino clustering (which can be important for models beyond the $\Lambda$CDM model, e.g. \cite{Baldi2014}).

However, it is not for all applications necessary to compute the sub-leading effects of non-linear clustering. Hence, in the recent years, a vast amount of attempts have been made to accurately compute the impact of massive neutrinos without the need for neutrino particles in the simulation. Since neutrinos in the observationally allowed mass range are well described in linear theory, one strategy involves realising the linear neutrino perturbations linearly on a 3D grid \cite{Brandbyge_2009, Brandbyge_2016, Tram_2019}. 

There are also several hybrid methods, that combine the advantages of grid and particle methods \cite{Bird_2018, Brandbyge_2010} or that include the back-reaction of the cold dark matter field on the evolution of neutrino perturbations \cite{Ali2012, Ali_Ha_moud_2012, Chen2021} trough a linear response approach. Alternatively, in \cite{Dakin2019}, the first three moments of the Boltzmann hierarchy are evolved non-linearly. Furthermore, there are even attempts to solve the 6D Vlasov-Poisson equation directly \cite{Yoshikawa2020}. A very recent development was done in \cite{Bayer_2021} where they successfully realised neutrinos as particles in the quasi-N-body code FastPM. Furthermore, there are several approximative methods, such the cosmological rescaling method \cite{Angulo2010, Zennaro_2019}. For a more complete review of simulation techniques, we refer the reader to \cite{angulo2021largescale}.

The method we are presenting here consistently includes the impact of the linear neutrino evolution on the non-linear clustering of matter, similar to e.g. \cite{Tram_2019}. However, based on a relativistic formulation of the problem \cite{Chisari2011} and extending previous work \cite{Partmann_2020}, we avoid any modifications to the N-body simulation beyond the Friedmann equation.

\subsection{A minimal method for the inclusion of neutrinos in N-body simulations}

In the following we outline our approach for the inclusion of massive neutrinos in standard Newtonian N-body simulations, without providing the technical details, which are contained in the following section~\ref{sec:Nmgauge}.
As discussed above, the impact of massive neutrinos on structure formation is mostly indirect, i.e. massive neutrinos do not significantly contribute to the small-scale matter overdensities due to their large thermal velocities preventing the gravitational collapse. Instead of directly participating in the growth of structures, the mere presence of massive neutrinos does affect the rate of structure formation of the cold species (in the following shortened to CDM+b). This motivates the use of a non-linear N-body simulations for CDM+b only, while the evolution of neutrinos is described by a linear Einstein-Boltzmann solver. 
  
In an ordinary Newtonian N-body simulation of cold matter, each particle is accelerated by the Newtonian potential $\Phi^{\N}$, and the particles thus follow a trajectory determined by 
\begin{equation}
 \label{eq:Newton}
 \partial_{\tau}^2 \boldsymbol{x}^{\N} + 2\mathcal{H}\partial_{\tau} \boldsymbol{x}^{\N} = - \nabla \Phi^{\N}(\boldsymbol{x}^{\N})  \, .
\end{equation}
The second term emerges from the expansion of the Universe with the conformal Hubble rate $\mathcal{H}$, that counteracts the gravitational collapse, while the last term is the usual Newtonian gravitational acceleration. 

In the presence of other particle species, such as massive neutrinos, the gravitational potentials are not only sourced by the cold massive particles, but receive important corrections from those additional species. Furthermore, in general relativity, particles follow geodesics which no longer need to coincide with the Newtonian trajectories. This is an effect that is usually neglected, but becomes important for large simulation volumes.

To capture these effects, we may still compute the motion of the massive particles using Eq.~\eqref{eq:Newton}, but we need to introduce an additional force term $\delta \boldsymbol{F}(\boldsymbol{x},t) $, which makes up for all the relativistic effects and the impact of the non-cold species. We can therefore generalise Eq.~\eqref{eq:Newton} to
\begin{equation}
\partial_{\tau}^2 \boldsymbol{x} + 2\mathcal{H}\partial_{\tau} \boldsymbol{x} = -\nabla \Phi^{\N}(\boldsymbol{x}) + \delta \boldsymbol{F}(\boldsymbol{x},t) \,,
\label{eqn:with_force}
\end{equation}
where the computation of ${\delta \boldsymbol{F}(\boldsymbol{x},t)}$ is a challenging problem that requires knowledge of full general relativity and the physics of all involved species. Since this additional force only depends on perturbatively small quantities, we can obtain a fully relativistic calculation of ${\delta \boldsymbol{F}(\boldsymbol{x},t)}$ using the \CLASS{} code \cite{Lesgourgues_2011}. In a Newtonian simulation, a realisation of this term can then be treated as an additional force \cite{Tram_2019, Brandbyge_2016}.

However, in order to minimise the modifications to existing N-body codes, we choose instead to perform a coordinate transformation, that absorbs the effect of $\delta \boldsymbol{F}(\boldsymbol{x},t) $ into a deformation of the coordinate system. This coordinate transformation $\boldsymbol{L}$ is defined by the condition, that in the new coordinate system $\boldsymbol{x}^{\mathrm{Nm}}=\boldsymbol{x}+\boldsymbol{L}$, the trajectories of matter particles (Eq.~\eqref{eqn:with_force}) become Newtonian

\begin{equation}
\partial_{\tau}^{2} \boldsymbol{x}^{\mathrm{Nm}}+2 \mathcal{H} \partial_{\tau} \boldsymbol{x}^{\mathrm{Nm}}=-\nabla \Phi^{\mathrm{N}}\left(\boldsymbol{x}^{\mathrm{Nm}}\right) \, .
\label{eqn:new_coords}
\end{equation}

The crucial step then is to find the coordinate shift $\boldsymbol{L}$ that relates the conventional coordinates (Eq.~\eqref{eqn:with_force}) to the tailor-made Newtonian motion coordinates in Eq.~\eqref{eqn:new_coords}.

In the following section \ref{sec:Nmgauge} we present the theoretical foundation of our method using the language of relativistic gauge theory. We then present our implementation in \GADGET{} \cite{Springel_2021} and \gevolution{} \cite{Adamek_2016} in chapter \ref{sec:Results}. We have written the manuscript such that the majority of chapter \ref{sec:Results} is accessible without the full theoretical justification of the method presented in section \ref{sec:Nmgauge}, which the reader may thus skip. We conclude in section \ref{sec:Conclude}. Finally we provide some numerical considerations in Appendix \ref{sec:compHT} and \ref{altImp} that may be relevant for readers interested to utilise our method or to reproduce our results.

\section{Newtonian motion gauges for massive neutrinos}
\label{sec:Nmgauge}
In this section, we present the mathematical method to interpret $\delta \boldsymbol{F}( \boldsymbol{x},t)$ as a deformation of the underlying coordinate system, such that the N-body code remains entirely agnostic of the relativistic effects. This approach is not necessarily based in general relativity and could be understood similar to the coordinate differences between a Eulerian or Lagrangian description of motion. It however fits seamlessly into the language of general relativity as a gauge choice. Deforming the coordinates to represent the extra forces is identical to choosing a gauge with a set of coordinates $\boldsymbol{x}^{\Nm}$ in which ${\delta \boldsymbol{F} (\boldsymbol{x}^{\Nm})=0}$.
We call such a gauge a Newtonian motion gauge (Nm gauge) since in this particular coordinate system the matter particles have exactly the same trajectories as the Newtonian particles in an ordinary Newtonian N-body simulation. On the other hand, all relativistic effects are expressed in the dynamic nature of the coordinate system on which these Newtonian trajectories are to be interpreted. 

In \cite{Fidler_2016} we have shown that there exists a class of gauges fulfilling these properties and that the underlying coordinates are well behaved and can be understood using linear (or weak field \cite{Fidler_2017_2}) perturbation theory. This framework therefore allows us to merge a linear Boltzmann code (such as \CLASS{}) and a non-linear N-body simulation (such as \GADGET{}) to obtain a fully relativistic and non-linear output. 
We have further shown that this framework can be generalised to include massive neutrinos \cite{Fidler:2018bkg,Partmann_2020}, making neutrino simulations possible using ordinary cold dark matter N-body simulations under the assumption that the simulation is started from a specific set of initial conditions and that the results are correctly interpreted. 
In this work, we aim to go beyond our previous results and make massive neutrino simulations simpler by using standard initial conditions and reducing the need to post-process the output, effectively making massive neutrino simulations as simple and efficient as cold dark matter simulations.

\subsection{Notation}

Before we start with the computations, we briefly summarise our adopted conventions.
We define a homogeneous and isotropic background with a perturbed metric by

\begin{align}
\label{eq:metric}
  \begin{split}
	g_{00} &= -a^2 (1+2A) \,, \\ 
	g_{0i} &= -a^2 \hat{\nabla}_i B \,, \\
	g_{ij} &= a^2 \left[ \delta_{ij} \left( 1 +2 \HL \right) - 2 \left( \hat{\nabla}_i\hat{\nabla}_j + \frac{\delta_{ij}}{3} \right) \HT \right] \,,
  \end{split}
\end{align}

where $A$ is the lapse perturbation, $B_i$ the shift vector, $\HL$ and $\HT$ the trace and the trace-free part of the spatial metric. The scale factor $a$ evolves with the Friedmann-Lemaître equation. We chose to use the conformal time $\tau$ with ${ad\tau = dt}$ as the time variable, and the derivatives with respect to the conformal time are denoted with a dot. The differential operator $\hat{\nabla}_i$ denotes the normalised gradient operator ${\hat{\nabla}_i= -(-\nabla^2)^{-1/2}\nabla_i}$ which is equal to ${-ik_i/|\boldsymbol{k}|}$ in Fourier space and $\delta_{ij}$ is the Kronecker symbol.

The stress energy tensor $T_{ij}$ is given by
\begin{align}
	T^0_0 &= - \rho \,, \\
	T_i^0 &= (\rho+p) \hat{\nabla}_i (v-B) \,,\\ 
	T^i_j &= p \delta^i_j + \left( \hat{\nabla}^i\hat{\nabla}_j +\frac{\delta_{ij}}{3} \right) \Sigma + (\rho + p)  \hat{\nabla}^i v \hat{\nabla}_j v  \,,
\label{Tmunu}
\end{align}
with the energy density $\rho$, the pressure $p$, the scalar velocity $v$ and the scalar anisotropic stress $\Sigma$.
Furthermore we introduce the two gauge invariant Bardeen potentials $\Psi$ and $\Phi$ as well as the comoving curvature perturbation $\zeta$, which will come in handy in various calculations
\begin{align}
\label{def:psi}
\Psi &\equiv A+ {\cal H} k^{-1} \left( B- k^{-1} \HTp \right) +k^{-1} \left( \dot{B}- k^{-1} \HTpp \right) \, ,
\\
\label{def:phi}
 \Phi &\equiv \HL +  \frac 1 3 \HT + {\cal H} k^{-1} \left( B- k^{-1} \HTp \right) \, ,
\\
\label{def:zeta}
\zeta &\equiv \HL + \frac 1 3 \HT + \mathcal{H} k^{-1}(B-v) \, .
\end{align}
The metric defined in Eq.~\eqref{eq:metric} is not fully specified and we can choose to enforce two gauge constraints. Common choices are the Poisson gauge, in which we enforce ${kB=\HTp}$ (temporal condition) and ${\HT=0}$ (spatial condition). Another important gauge is the N-boisson gauge that shares the time coordinates with the Poisson gauge but sets ${\HT = 3\zeta}$. We have shown in \cite{Fidler_2015,Fidler_2017_2} that this gauge is consistent with ordinary Newtonian N-body simulations in a radiation-free\footnote{In this work, radiation-free implies that no relativistic species (photons or massive/massless neutrinos) are present, while cold species (CDM and baryons) and a cosmological constant are allowed.} Universe in the sense that the computed Newtonian trajectories agree with the relativistic motion, in agreement with \cite{Chisari2011}. Those coordinates are therefore always implicitly assumed when an ordinary Newtonian N-body simulation is analysed. 
In cosmologies that include relativistic species (radiation) there are still gauges with Newtonian trajectories, but their metric potentials usually need to be determined numerically. We label those as the Newtonian motion gauges. Note that all those gauges only differ in their choice of spatial gauge condition while the temporal gauge choice is always kept identical to the Poisson gauge \cite{Fidler_2017_2}.
Therefore the fractional overdensity ${\delta = \frac{\delta\rho}{\bar\rho}}$ is identical for any Newtonian motion gauge: ${\delta^{\Nm} = \delta^{\rm P}}$ with `Nm' referring to Newtonian motion gauge and `P' to the Poisson gauge\footnote{The fractional overdensity $\delta$ is only dependent on the temporal gauge condition.}. 

In addition to the relativistic densities labeled by their gauge, we also define the gauge agnostic simulation densities $\delta^{\N}$. Those refer to the density as measured from a Newtonian N-body simulation assuming a Euclidean geometry. The geometrical volume deformation establishes the relation
\begin{equation}
\delta^{\N} = \delta^{\Nm}_{\rm m} + 3 \HL \, ,
\end{equation}
which for the N-boisson gauge in a radiation-free Universe simplifies to 
\begin{equation}
\delta^{\N} = \delta^{\rm P}_{\rm m} - 3 \zeta + 3 \Phi \approx \delta^{\rm syn} \, ,
\end{equation}
confirming that the densities measured from ordinary N-body simulations are compatible with the relativistic synchronous gauge densities. 

\subsection{Backscaling initial conditions}

The most commonly employed method to generate initial conditions for Newtonian N-body simulation is called backscaling. Here, we outline the general method, and how it is usually applied, before elaborating on our new method. In common backscaling, one obtains the initial conditions by computing the (relativistic) linear matter power spectrum at the final time, and scales it back in time using the linearised Newtonian growth equation
\begin{equation}
\label{eqn:newtonian_growth_ode}
(\partial_{\tau} + \mathcal{H}) \dot{\delta}^{\N} - 4\pi G a^2 \rho_{\rm m} \delta^{\N} = 0\,.
\end{equation}
Being a second order differential equation, two linear independent solutions exist which we label as the growing mode $D_+$ and decaying mode $D_-$, allowing us to represent any initial density and velocity by a corresponding mixture of growing and decaying modes. Since the decaying mode quickly looses importance relative to the growing mode, we can, at sufficiently late times, approximate the solution by a pure growing mode. 
In that case, the solution of Eq.~\eqref{eqn:newtonian_growth_ode} simply reads
\begin{equation}
\label{eqn:backscaling}
\delta^{\N}(\tau_i) = D_+(\tau_i)\delta^{\N}(\tau_f)\,,
\end{equation}
with the initial time $\tau_i$ and the final time $\tau_f$.
Eliminating the decaying mode, the velocity is now uniquely tied to the density via
\begin{equation}
kv^{\N} = -\dot{\delta}^{\N} = -\dot{D}_+ \delta^{\N} \, .
\end{equation}

In backscaling, the linear final time matter power spectrum obtained by an Einstein-Boltzmann solver is used to obtain an equivalent version at early time assuming a pure growing mode by employing Eq.~\eqref{eqn:backscaling}. Initial particle velocities are set assuming their growing mode values based on the densities. This method creates an artificial matter-only Universe at initialisation which emulates our real Universe on large scales and by design gives the desired relativistic output, at least at the final time. This way, we can introduce relativistic effects on linear scales, and let the simulation figure out the dynamics on the small scales. Whereas this method seems quite simple, and also easy to use, it is not suitable for every cosmology. In particular, backscaling breaks down in a massive neutrino cosmology, or at least requires some non-trivial modifications \cite{Fidler_2019_2}.

\subsection{Backwards Newtonian motion gauges}
\label{sec:bwNmgauge}
In addition to massive neutrinos complicating the procedure, backscaling is not usually formulated in relativistic theory and it is not apparent which gauge should be used for the present-day power spectrum. In \cite{Fidler_2017} we have answered this question and shown that the backscaling method can be consistently embedded in a certain type of Newtonian motion gauges, that we call backwards Newtonian motion gauges.

We now develop the concept of a backwards Newtonian motion gauge for massive neutrinos cosmologies. In \cite{Fidler_2019_2} we found that backscaling the entire matter power spectrum leads to significant problems. Therefore, in this work we only aim to exploit backscaling for CDM+b and do not realise neutrino particles in the N-body simulation. We will further examine if the synchronous gauge power spectrum is still suitable for backscaling, even when considering massive neutrinos.

As in all Newtonian motion gauges, we must obey the Newtonian motion gauge condition \cite{Fidler_2016} to succeed in absorbing the extra forces and achieving a Newtonian motion for CDM+b
\begin{equation}
\label{eqn:H_T_ode}
(\partial_{\tau} + \mathcal{H}) \HTp - 4\pi G a^2 \rho_{\rm m} (\HT - 3\zeta) = S \, ,
\end{equation}
where $S$ is given by
\begin{equation}
S = 4 \pi G a^2 \rho_{\rm other} \delta_{\rm other} + 3 \mathcal{H} k^{-1}(\rho+ p)_{\rm other} (v-B) + 2 p \Pi \, ,
\end{equation}
and where the index `m' refers to the particles in the simulation (cold dark matter and baryons) only, while `other' refers to all species that are present in the Universe beyond cold dark matter and baryons and are thus not realised as particles in the simulation and thereby not included in the Newtonian gravitational potential.

The differential equation for $\HT$, describing the dynamic deformation of the coordinate system, does not completely fix the gauge conditions. A two-parameter residual freedom remains that allows for an identification of the simulation density and velocity at one given point in time. According to the geometric relation of volume deformation induced by $\HL$ we have
\begin{equation}
\label{eqn:ic_in_nm}
\delta^{\N}(\tau_i) = \delta^{\Nm}_{\rm m}(\tau_i) + 3 \HL(\tau_i) = \delta^{\rm P}_{\rm m}(\tau_i) - \HT(\tau_i) + 3 \Phi(\tau_i) \,,
\end{equation}
while for the velocities we have
\begin{equation}
\label{eqn:ic_in_nm_velocity}
v^{\N}(\tau_i) = v^{\Nm}_{\rm m}(\tau_i) + k^{-1} \HTp(\tau_i)\,.
\end{equation}
This allows us to control both the simulation density and velocity via initial values for $\HT$ and $\HTp$.
In a forwards Nm gauge, we usually fix our gauge to the N-boisson gauge at initial time, i.e. $\HT(\tau_i)=3\zeta(\tau_i)$ and $\HTp(\tau_i) = 3\dot{\zeta}(\tau_i)$.
In backscaling however, we utilise the present time power spectrum to initialise the simulation which makes it natural to fix the residual gauge freedom at the final time and solve the ODE backwards. This defines the term of a backwards Newtonian motion gauge and 
in \cite{Fidler_2019_2} we have concluded that we can set $\HT(\tau_f) = 3\zeta(\tau_f)$ and the derivative $\HTp(\tau_f) = 3\dot{\zeta}(\tau_f)$ such that the gauge formally agrees with the N-boisson gauge at the final time:
\begin{equation}
\delta^{\N}(\tau_f) =  \delta^{\rm P}_{\rm m}(\tau_f) - \HT(\tau_f) + 3 \Phi(\tau_f) =  \delta^{\rm P}_{\rm m}(\tau_f) - 3 \zeta(\tau_f) + 3 \Phi(\tau_f) \,.
\end{equation}

\paragraph{Eliminating the decaying modes}
We notice that there is a more optimal choice for the gauge fixing relevant especially in massive neutrino cosmologies. While we aim to have $\HT=3\zeta$ at the final time to make the particle positions comparable to ordinary radiation-free simulations evaluated in the N-boisson gauge, the derivative $\HTp$ can be varied more freely. Indeed, if we want to make sure that the Newtonian density is a pure growing mode as assumed by backscaling, we find
\begin{equation}
\label{eqn:H_T_boundary}
\dot{\delta}^{\N} =  \dot{\delta}^{\rm P}_{\rm m}(\tau_f) - \HTp(\tau_f) + 3 \dot{\Phi}(\tau_f) = \dot{D}_+ \left[\delta^{\rm P}_{\rm m}(\tau_f) - \HT(\tau_f) + 3 \Phi(\tau_f) \right] = \dot{D}_+ \delta^{\N} \, ,
\end{equation}
that can alternatively be expressed as
\begin{equation}
\HTp(\tau_f) = \theta_{\rm m}^{\N} - \theta_{\rm m}^{\rm P} \, ,
\end{equation}
using the velocity divergences $\theta$.

Together with $\HT(\tau_f) = 3\zeta$, this condition completely fixes the gauge and ensures that the Newtonian density evolves as a pure growing mode, opposed to only being close to a growing mode in \cite{Fidler_2017}. This means that we can obtain our initial conditions exactly by backscaling and may utilise well established methods such as 2LPT or higher order initial conditions that are based on a Newtonian growing mode.

We now obtain valid initial conditions by backscaling the present-day power spectrum of  
\begin{equation}
\label{eqn:delta_N_present}
\delta^{\N}(\tau_f) =  \delta^{\rm P}_{\rm m}(\tau_f) - 3 \zeta(\tau_f) + 3 \Phi(\tau_f) \approx \delta_{\rm m}^{\rm syn}(\tau_f) \, ,
\end{equation}
where $\delta_{\rm m}^{\rm P}$ is the Poisson gauge overdensity of cold dark matter and baryons only. We find that this combination is almost identical to the synchronous gauge density also in the case of a massive neutrino cosmology. Only on scales larger than $k < 10^{-3} \si{.Mpc}^{-1}$ we find permille level deviations. Using backscaling based on the synchronous gauge therefore remains a suitable choice.

\paragraph{Solving for $\HT$}
Following the same reasoning as in \cite{Fidler_2017}, we will now derive the exact form of the solution for $\HT$. We find that the Newtonian growth Eq.~\eqref{eqn:newtonian_growth_ode} and the ODE for the metric potential $H_{\rm T}$ \eqref{eqn:H_T_ode} share a set of homogenous solutions. Therefore we can formally express both using the dark matter growing and decaying modes
\begin{equation}
\label{eqn:H_T_ansatz}
\HT(\tau) = C_+(\tau) D_+ (\tau ) + C_-(\tau)D_-(\tau) + 3 \zeta(\tau) \, ,
\end{equation}
where $C_+$ and $C_-$ are two time dependent functions that need to be determined and depend on the boundary conditions of the ODE and the inhomogeneous part of eq.~\eqref{eqn:H_T_ode}. The last term $3\zeta$ is an analytically known part of the special solution of the equation, that we add explicitly such that in a pure dark matter case the coefficients $C_\pm$ vanish trivially and we reproduce the N-boisson gauge.  

\begin{figure}[htpb]
\centering
  \includegraphics[width=0.9\linewidth]{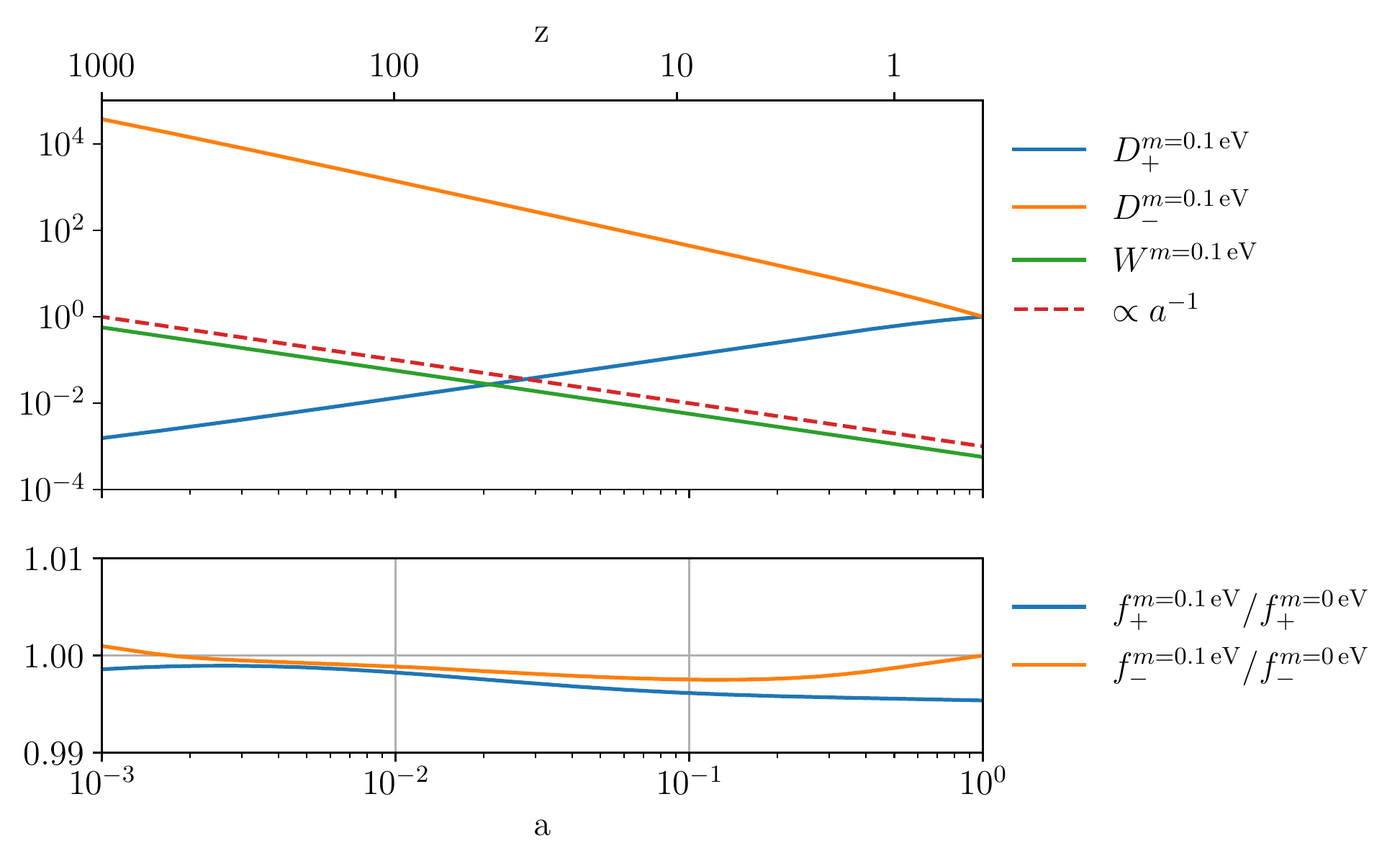}
 \caption{The upper plot shows the growing mode $D_+$ and decaying mode $D_-$ in a massive neutrino cosmology. The Wronskian is inversely proportional to the scale factor $a$. The lower plot shows the suppression of the growth rate $f_+$ compared to a massless neutrino cosmology.}
  \label{fig:modes}
\end{figure}

The growing mode $D_+$ can be obtained fairly straightforward by numerical forwards integration for example using an Einstein-Boltzmann code, such as \CLASS{}. To find the decaying mode $D_-$, we make use of the Wronskian
\begin{equation}
 W = D_+\dot{D}_- + \dot{D}_+D_- ,
\end{equation} 
which has a simple equation of motion $\dot{W} = -\mathcal{H}W $ and is analytically solved by being proportional to $a^{-1}$. Using $W$, we then obtain a first order ODE for $D_-$ which can easily be solved by integration. Figure~\ref{fig:modes} shows the growing and decaying modes for a massless and a massive neutrino cosmology highlighting that massive neutrinos affect the rate of structure formation for CDM+b via their impact on the Hubble rate. In the lower plot we show the growth rate as the logarithmic derivative of the growing mode $f_+ = d \ln D_+ / d \ln a$, which is suppressed in the massive neutrino cosmology.

Following the method of the variation of the constants to solve Eq.~\eqref{eqn:H_T_ansatz}, we use the fact that we have two degrees of freedom, $C_-$ and $C_+$ but only one constraint. This gives us the freedom to choose one supplementary equation. Note that this is an arbitrary choice that has no impact on the obtained solution. However, in order to enforce the boundary conditions that we derived in \eqref{eqn:H_T_boundary}, a convenient option is to require 
\begin{equation}
\label{eqn:H_T_extra_condition}
\dot{C}_+D_+ + \dot{C}_-D_- + 3\dot{\zeta} = \HTp(\tau_f) \, ,
\end{equation}
where the value of $\HTp$ is given by Eq.~\eqref{eqn:H_T_boundary}. By plugging the ansatz into the ODE and making use of the supplementary condition Eq.~\eqref{eqn:H_T_extra_condition}, we arrive to the following equation:
\begin{equation}
\label{eqn:H_T_second_equation}
\dot{C}_+ \dot{D}_+ + \dot{C}_-\dot{D}_-  =S -  \mathcal{H} \HTp(\tau_f) \,.
\end{equation}
Now we have two equations (Eq.~\eqref{eqn:H_T_extra_condition} and Eq.~\eqref{eqn:H_T_second_equation}) for two degrees of freedom, which are solved by the semi-analytical solution  
\begin{equation}
\label{eqn:C_plus_minus}
 C_\pm(\tau) = \pm  \int_{\tau}^{\tau_f} \left( D_\pm(\tilde{\tau}) S_-(\tilde{\tau}) - \dot{D}_\pm(\tilde{\tau})S_+(\tilde{\tau}) \right) W^{-1}(\tilde{\tau}) d\tilde{\tau} \,,
\end{equation}
where $S_+(\tau)=\HTp(\tau_f) -3\dot{\zeta}(\tau)$ and $S_-(\tau) = S(\tau) - \mathcal{H} \HTp(\tau_f) $.
In this solution $\HT$ does not only get a contribution from radiation and massive neutrinos, but also takes into account that the simulation is representing a pure growing mode Universe.

\begin{figure}[htpb]
\centering
  \includegraphics[width=0.9\linewidth]{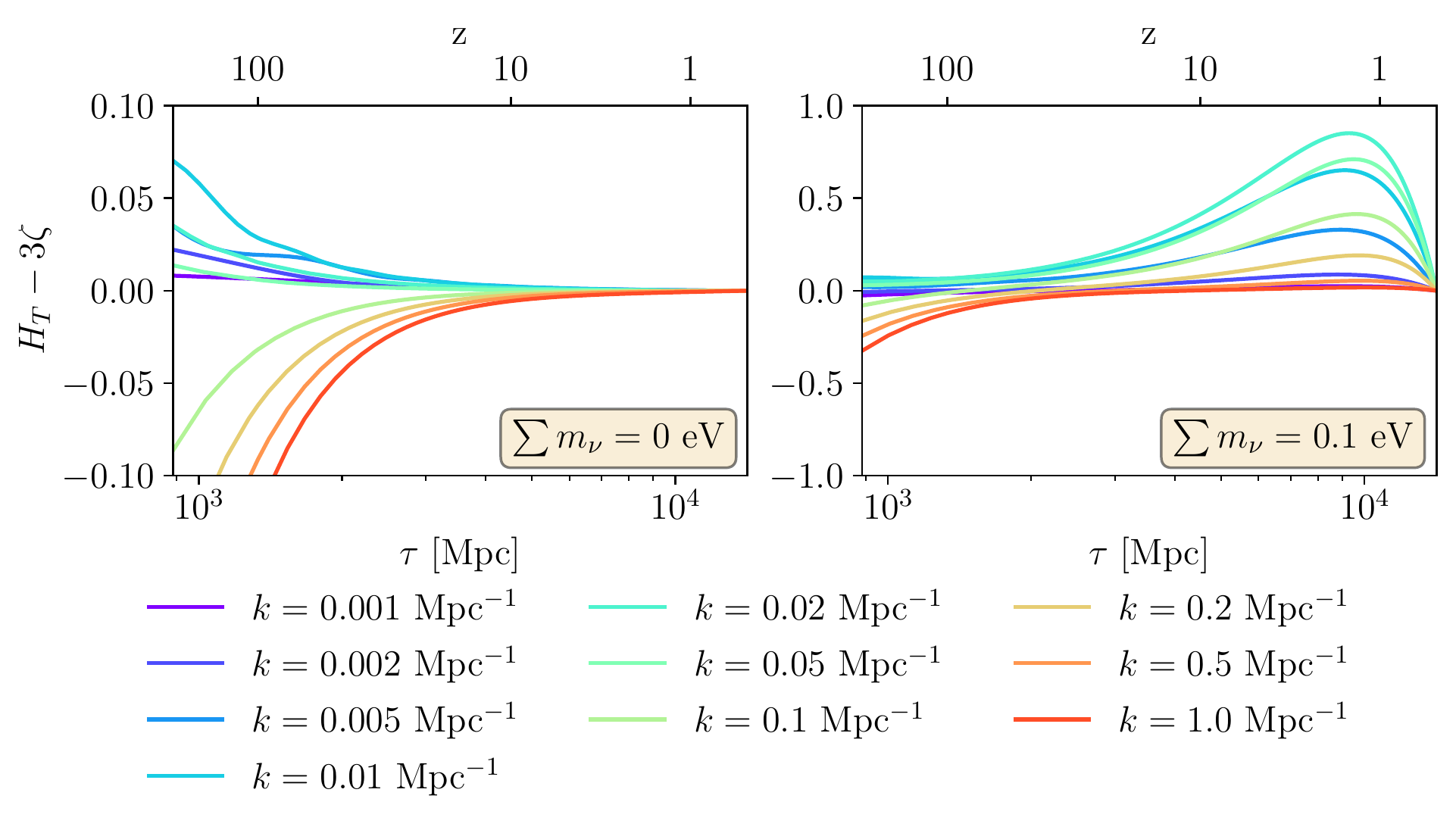}
 \caption{Shown is the deviation of $\HT$ from $3\zeta$ as a function of the conformal time. In a radiation-free case (left plot), the deviation remains close to zero during the entire late Universe since the importance of radiation is small and the solution stays close to the N-boisson gauge. In the massive neutrino cosmology (right plot) $\HT$ is equal to $3\zeta$ at the final time by construction. The impact of the massive neutrinos then quickly induces a significant difference. The modes are normalised to $\zeta=-1$ on super-horizon scales following the \CLASS{}-convention.}
  \label{fig:H_T_tau}
\end{figure}

We employ $\HT - 3\zeta$ as an indicator of how much the coordinates have been deformed compared to the N-boisson gauge ($\HT = 3\zeta$). As ordinary N-body simulations in a radiation free Universe are analysed in the N-boisson gauge, this provides a measure for the amount of corrections absorbed in our choice of coordinates. A very small value of $\HT - 3\zeta$ indicates that the coordinates are close to the N-boisson gauge and thus the output of the simulation is directly comparable to that of a massless neutrino cosmology without the need of any post-processing. Figure \ref{fig:H_T_tau} shows $\HT - 3\zeta$ as a function of time. One can easily see that $\HT$ coincides with $3\zeta$ at $z=0$, which is one initial condition of our ODE. The second feature is the non-vanishing slope at the final time, which produces a little bump around $z=1$. Our result differs from the previously presented backwards method in \cite{Fidler_2019_2}, which has a different slope at the final time and relies on initialising the simulation with specific initial conditions. Our new solution combines the ease to use common backscaling initial conditions with generally small values for $\HT$, implying that the particle positions updates are small.

Note that even in the case that $\HT - 3 \zeta$ remains negligible, the simulation already contains the impact of massive neutrinos, via the modified Hubble rate and the initial conditions. Both changes, compared to a pure CDM case, do effect the simulation fully non-linearly. The metric potential $\HT$ only describes the residual difference between the full relativistic theory and the simulation that already contains those important effects. 

For practical purposes, we present a more efficient numerical method to construct $\HT$ in Appendix \ref{sec:compHT} using tools that are available directly in the \CLASS{} code. In our work we have implemented both approaches to crosscheck our solutions.

\section{Results}
\label{sec:Results}

The theoretical framework presented in the previous chapter provides a consistent prescription for N-body simulations with massive neutrinos. This prescription is derived from first principles and allows to run accurate N-body simulations with massive neutrinos with only minor modifications to the underlying N-body code. Our novel method is based on three simple steps:

\subsection*{Step 1: Friedmann equation}

The most important effect of massive neutrinos appears on the level of the background evolution, i.e. in the Friedman equation. Because changing the expansion rate of the Universe does affect all particle species (dark matter, baryons, photons) the additional contribution from the mass of neutrinos affects the overall clustering rate of matter:

\begin{equation}
\left[H(a) / H_{0}\right]^{2}=\Omega_{\mathrm{m} 0} a^{-3}+\Omega_{\nu}(a)+\Omega_{\gamma 0} a^{-4}+\Omega_{\Lambda 0} \,.
\end{equation}

The computation of $\Omega_{\nu}(a)$ is straightforward and only requires the integration of the neutrino phase-space at the background level. In N-body simulations with comoving coordinates, the scale factor appears on the level of the time-integration, i.e. in the kick and drift operations that are applied to the N-body particles. As we argued, this change in the expansion history is one of the reasons for the suppression of the matter power spectrum on small scales, if the total matter energy density at final time is kept fixed.

\subsection*{Step 2: Backscaling initial conditions}

As derived in chapter \ref{sec:bwNmgauge}, the goal of the method is to absorb the neutrino effects beyond the background evolution into the initial conditions of the simulation. We have shown that this approach is consistent with the commonly employed backscaling initial conditions. The initial simulation density is gained by scaling the final CDM+b power spectrum back in time, i.e. 

\begin{equation}
\delta^{\N}(z_{i}) =  D_+(z_{i}) \, \delta^{\N}(z_f) \, .
\label{backscaling}
\end{equation}

Let us stretch that we use only CDM+b and not the total matter spectrum. Furthermore, the initial power spectrum, which can be computed with a linear Boltzmann code, must be provided in the correct general relativistic gauge. It can either be constructed from the Poisson gauge quantities (see Eq.~\eqref{eqn:delta_N_present}), or to high accuracy it can be approximated by the synchronous gauge matter power spectrum. It is also important to use the correct $D_+$ in Eq.~\eqref{backscaling}, which is unique for each cosmology.

Crucially, the initial conditions are constructed such that they only contain a growing mode. In this case, the relation between the overdensity $\delta^{\N}$ and its time derivative is determined by the scale independent linear growth factor $D_+$ by

\begin{equation}
\label{eqn:H_T_boundary2}
\dot{\delta}^{\N} = \dot{D}_+ \delta^{\N} \, .
\end{equation}

Because $\delta^{\N}$ and $\dot{\delta}^{\N}$ determine the initial particle positions and velocities respectively, the particle displacements and initial velocities are now linearly dependent on all scales. This is different from the case presented in \cite{Partmann_2020}, where the initial particle positions and velocities were independent and had to be initialised from two separate transfer functions for velocities and positions. 

Furthermore, $\delta^{\N}$ is constructed such that it evolves according to the Newtonian equations of motion. Hence, it is now possible to use an ordinary initial conditions generator that employs either the Zel'dovich approximation or higher order Lagrangian perturbation theory (LPT) as presented in \cite{Bouchet_1994,Jenkins_2010}. In second order LPT, if $\boldsymbol{q}$ denotes the particle positions in a homogeneous template and $\phi^{(i)}$ the first and second order potentials, the initial particle displacements are given by 

\begin{eqnarray}
\label{LTP_pos}
\mathbf{x} =& \mathbf{q}-D_{1} \nabla_{q} \phi^{(1)}+D_{2} \nabla_{q} \phi^{(2)} \, , \\ 
\label{LTP_vel}
\mathbf{v} =& -D_{1} f_{1} H \nabla_{q} \phi^{(1)}+D_{2} f_{2} H \nabla_{q} \phi^{(2)} \, .
\end{eqnarray}

Here, $f_i$ is the logarithmic derivative of $D_i$, i.e. $f_i = d \ln D_i  / d \ln a$. In principle, the functions $D_1$, $D_2$ and their derivatives $f_1$, $f_2$ can directly be computed numerically as a solution to two different ordinary differential equations: The equation for $D_1$ is the Newtonian growth equation \eqref{eqn:newtonian_growth_ode}, with $D_1 = D_+$. The defining equation for $D_2$ is given by

\begin{equation}
(\partial_{\tau} + \mathcal{H}) \dot{D}_2 - 4 \pi G a^2 D_2 = - 4\pi G a^2 (D_1)^2 \, .
\end{equation} 

In many initial conditions generators, it is common practice to use approximations for $f_1$, $f_2$, $D_1$ and $D_2$. For example, \cite{Bouchet_1994} found $f_1 \approx [ \Omega_m(z)]^{5/9} $ and $f_2 \approx 2 [ \Omega(z) ]^{6/11}$. However, these approximations are only accurate in cosmologies without massive neutrinos in the Hubble rate. Therefore, we use the directly computed values for $f_1$, $f_2$, $D_1$ and $D_2$ instead.

\subsection*{Step 3: Particle displacements}
The particle positions obtained from our simulation are formally in the coordinates of the underlying Newtonian motion gauge. Often, an analysis in a more common gauge will be preferable. To this end, we can apply a post-processing shift to the particle positions to obtain their positions in the Poisson or N-boisson gauges. Here we focus on obtaining the output in the N-boisson gauge, making the further analysis equivalent to that of any Newtonian radiation-free simulation that also provides output in that gauge \cite{Adamek:2019aad}. 

\begin{figure}[t]
\centering
  \includegraphics[width=0.9\linewidth]{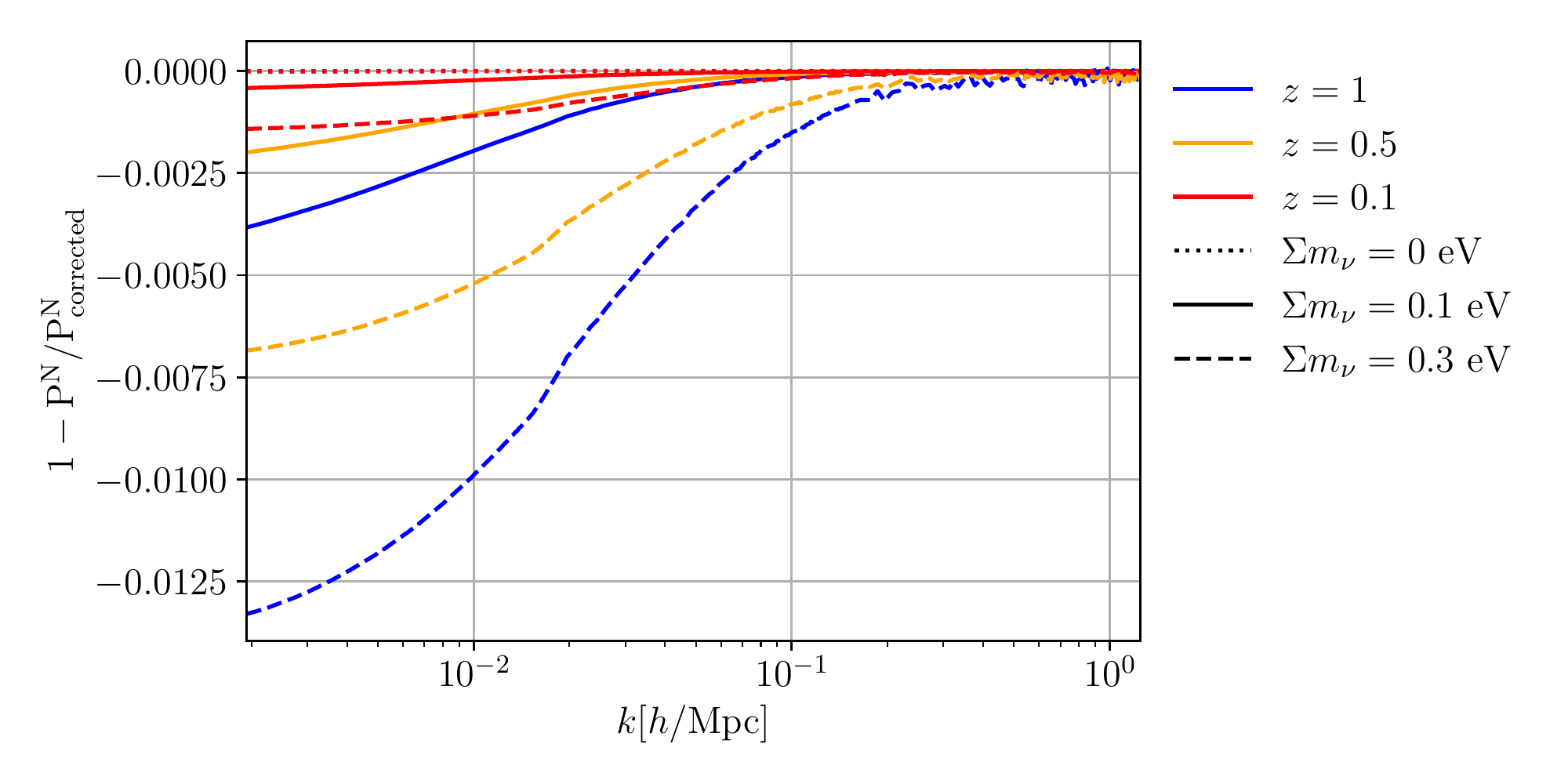}
 \caption{The ratios of the corrected and uncorrected power spectra in cosmologies with different neutrino masses. The initial conditions are constructed such that no correction is necessary at ${z = 0}$. For ${z \neq 0}$, the particles in a simulation snapshot must be displaced in order to be comparable with ordinary N-body simulations. However, for small $z \approx 0$ and small scales, this correction remains negligible.}
 
 \label{fig:different_z_different_masses}
\end{figure}

The shift of the particle positions can be implemented similarly to an initial conditions code using a displacement field, only that in this case the displacement is based on the value of $\HT - 3\zeta$ as discussed in section \ref{sec:bwNmgauge}. In figure \ref{fig:different_z_different_masses}, we show the ratio between the corrected and uncorrected power spectra at different redshifts. We see that for redshifts smaller than 1 and on small scales $k > 10^{-1} \si{.Mpc}^{-1}$, the two spectra agree to high precision. However, for higher neutrino masses, on the large scales and early times, the corrections grow to the percent level. Note that at $z=0$ the displacement vanishes by construction and no correction is required. In principle it is possible to optimise this choice for a given survey, see Appendix \ref{altImp} for more details.

\subsection{Simulations}
To validate our method, we implemented our backscaling prescription in two publicly available N-body codes \GADGET{} \cite{Springel_2021} and \gevolution{} \cite{Adamek_2016}. 

In \GADGET{}, we use the built-in initial conditions generator N-genic \cite{Springel2015NGenICCS}, that supports Lagrangian perturbation theory up to second order. As presented in the previous chapter, we modified the factors $f_1$, $f_2$, $D_1$ and $D_2$ according to \eqref{LTP_pos} and \eqref{LTP_vel} to include the impact of neutrinos on the linear growth and use the backscaled power spectrum that was generated with \CLASS \cite{Lesgourgues_2011}. We add the neutrino density to the background evolution and then use the tree-particle-mesh algorithm to evolve the initial conditions, which are set up at $z=100$, up to the final time $z = 0$. The comoving gravitational softening length is chosen to be $\mathcal{O}(1/40)$ of the mean inter-particle separation.

For simulations with \gevolution{}, we use the built-in initial conditions generator, that does not use LPT but instead initialises the particle positions directly from a linear density $\delta(k)$ and the velocity divergence ${\theta(k)}$. Hence, we use the transfer functions ${\delta^{\N}}$ and ${\theta^{\N} = -\dot{D}_+\delta^{\N}}$ as computed by \CLASS{} according to the backscaling scheme. Although \gevolution{} supports general relativistic corrections, we only use its particle mesh algorithm to solve the Newtonian Poisson equation.  

Since both codes use different prescriptions for the initial conditions, we have two independent implementations to validate our method. The numerical parameters of all simulations are summarised in table \ref{table:pars}. We investigate cosmologies with different neutrino masses, where the critical CDM density is always chosen such that the total matter content at ${z = 0}$ is equal to the standard radiation-free cosmology, i.e. ${\omega_{\rm m} = \omega_{\rm cdm} + \omega_{\rm b} + \sum m_{\nu} / 93.14 \si{.eV}}$ is kept fixed. The cosmological parameters are shown in table \ref{table:cosmo}.

\begin{table}[t]
\begin{center}
\begin{tabular}{c c c c c c c c}

Run & Code & $\sum m_\nu \, [\si{eV}]$ &$N_{\rm p}^{1/3}$ & $N_{\rm g}^{\rm PM}$ & $L \, [\si{Gpc}/h]$&$l_{\rm soft} \, [\si{Mpc}/h]$  & ICs \\ 
\hline
\hline
r1 & \gevolution{} &  0, 0.1, 0.2, 0.3 & 2048 & 2048 & 32.0 & -  & $\delta^{\rm N}$ and $\theta^{\rm N}$ \\  
r2 & \gevolution{} & 0, 0.1, 0.2, 0.3 & 2048 & 2048 & 8.0 & -  & $\delta^{\rm N}$ and $\theta^{\rm N}$ \\  
r3 & \gevolution{} &  0, 0.1, 0.2, 0.3& 2048 & 2048  & 2.0 & - &$\delta^{\rm N}$ and $\theta^{\rm N}$ \\  
r4 & \gevolution{} &  0, 0.1 & 2048 & 2048  & 1.0 & - &$\delta^{\rm N}$ and $\theta^{\rm N}$ \\  
r5 & \gevolution{}  &  0, 0.1 & 2048 & 2048 &  0.75 & - & $\delta^{\rm N}$ and $\theta^{\rm N}$ \\  
r6 & \gevolution{}  &  0, 0.1, 0.2, 0.3  & 2048 & 2048 & 0.5 & - &$\delta^{\rm N}$ and $\theta^{\rm N}$ \\  
\hline
\hline
r7 & \GADGET{} & 0, 0.1 & 512 & 2048 & 32.0& 1.6 & 2-LPT\\  
r8 & \GADGET{} & 0, 0.1 & 512 & 2048 & 8.0 & 0.4  & 2-LPT\\  
r9 & \GADGET{} & 0, 0.1 & 1024 & 2048 & 2.0 & 0.05  & 2-LPT \\  
r10 & \GADGET{} & 0, 0.1 & 1024 & 2048 & 1.0 & 0.025  & 1-LPT\\  
r11 & \GADGET{} & 0, 0.1 & 1024 & 2048 & 1.0 &  0.025  & 2-LPT\\  
r12 &\GADGET{} & 0, 0.1 & 1024 & 2048 & 0.5 & 0.0125  & 2-LPT\\  
\hline
\hline
\end{tabular}

\end{center}
\caption{\label{table:pars}Numerical setting used for simulations in \GADGET{} and \gevolution{}. $N_{\rm p}$ and $N_{\rm g}^{\rm PM}$ denote the number of particles and particle mesh grid cells per dimension, respectively. $L$ and $l_{\rm soft} $ are the co-moving box size and gravitational softening length. The initial particle positions and velocities with \gevolution{} are initialised from independent transfer functions $\delta^{\rm N}$ and $\theta^{\rm N}$. In \GADGET{}, we use standard 1-LPT or 2-LPT which only depends on the density transfer function $\delta^{\rm N}$.}
\end{table}

\begin{table}[t]
\begin{center}
\begin{tabular}{c c c c c c}
 $\sum m_{\nu}[\si{eV}] $  & $ m_1[\si{eV}]$ & $ m_2[\si{eV}]$  & $ m_3[\si{eV}]$  & $N_{\rm eff}$ & $\omega_{\rm cdm}$ \\
\hline
\hline 
 0 & 0 & 0 & 0 & 3.046 &  0.12038\\ 
 0.1 &  0.0219337  & 0.023582 & 0.054485 & 0.00641 & 0.119306  \\ 
 0.2 & 0.0604912   & 0.061108 & 0.078401 & 0.00641 & 0.118233 \\ 
 0.3 & 0.0958014 & 0.096192 & 0.108007 & 0.00641 & 0.117159 \\ 
\end{tabular}
\end{center}
\caption{\label{table:cosmo}Cosmological parameters used in our simulations. The chosen neutrino masses $m_i$ are compatible with the normal mass hierarchy. $N_{\rm eff}$ is the number of effective ultra-relativistic degrees of freedom, and  $\omega_{\rm cdm}$ denotes the CDM density parameter.}
\end{table}

\begin{figure}[t]
\centering
  \includegraphics[width=0.98\linewidth]{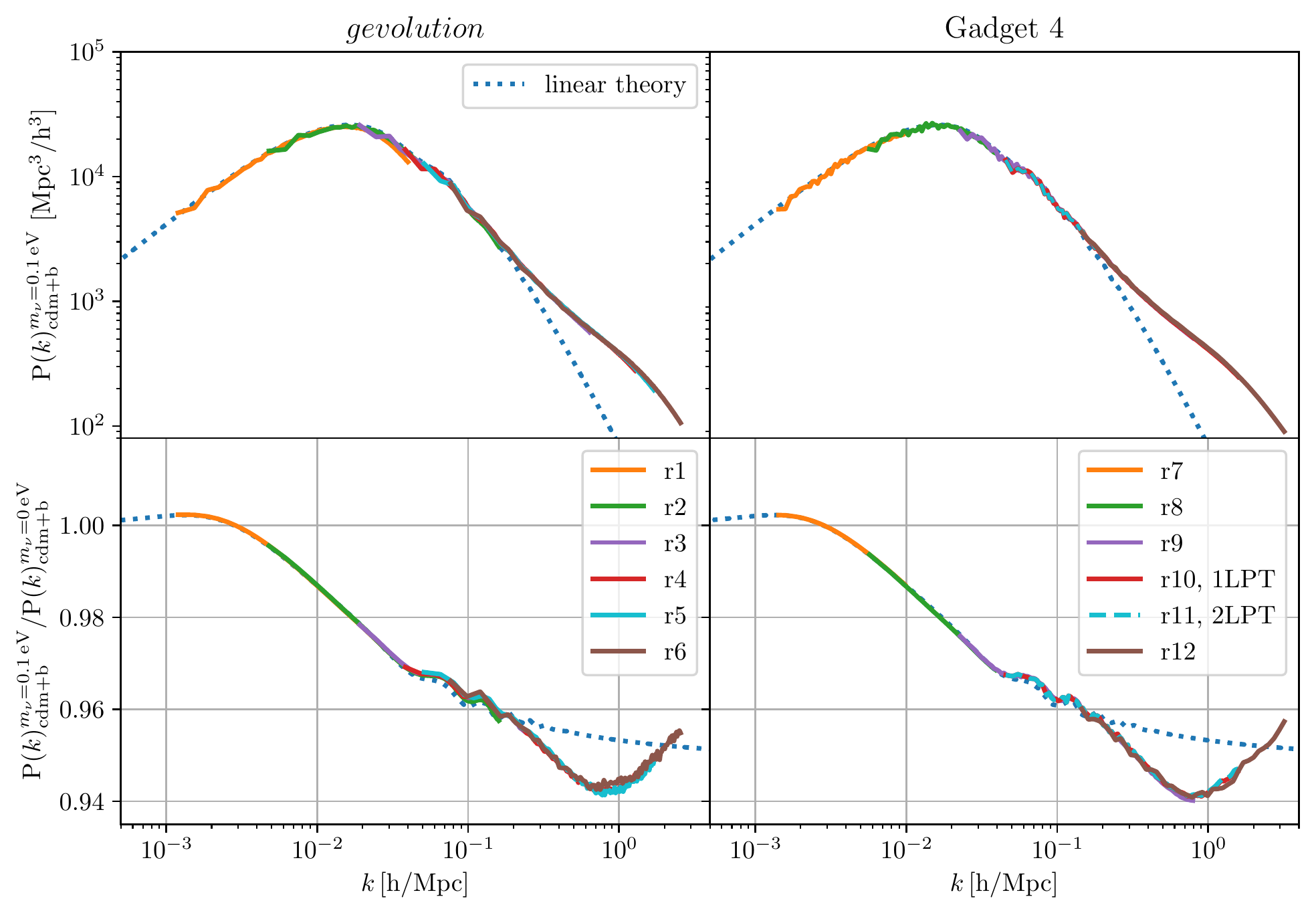}
 \caption{Final time (z=0) power spectra of our simulations with $\sum m_\nu = 0.1  \si{.eV}$ in \gevolution{} (left panel) and \GADGET{} (right panel) for different box sizes. The bottom panels show the comparison of the massive neutrino to our reference cosmology with massless neutrinos. For wave numbers larger than the Nyqvist frequency of the simulation, the relative power spectrum converges to the unique, spoon shaped, solution. Note that the \GADGET{} simulations have a smaller number of particles, but a higher force accuracy on small scales compared to \gevolution{}. Both codes and initial conditions methods produce similar results.}
  \label{fig:pk}
\end{figure}

The power spectra at final time for all simulations with $\sum m_\nu = 0.1  \si{.eV}$ are shown in the top panels of figure \ref{fig:pk}. On large scales, we recover the linear expectation (dashed curve), while the non-linear enhancement of the power spectrum is clearly visible on small scales. As argued in the previous chapter, for those results it was not necessary to displace the particles at final time - instead, we could directly take the snapshot from the simulation or measure the power spectrum on the fly. Note however that the shown spectra only contain CDM and baryon perturbations.

To prove the convergence of our results, we show several simulations with different box sizes $L_{\rm Box} $ and particle numbers $N_{\rm p}$. On wave numbers significantly smaller than the Nyqvist frequency $k_{\rm Nyqvist} = \pi N_{\rm g}/ L_{\rm Box}  $, the power spectra are converged to a resolution independent solution. Due to the TreePM algorithm in \GADGET{}, $N_{\rm p} = 1024^3$ particles are enough to reach a higher spatial resolution than \gevolution{} ($N_{\rm p} = 2048^3$), although the particle number is eight times smaller. 

Since neutrinos do not appear in the simulations directly, the plots essentially show a convergence study of the pure, Newtonian, dark matter simulation with the respective N-body codes. Hence, the small differences between the power spectra at different resolutions and box sizes are measure of convergence of the underlying CDM simulations, which is not related to the accuracy of the neutrino method. 

In the lower panels, we compare the resulting power spectrum of a massive neutrino cosmology to the reference cosmology with massless neutrinos. With both codes, we recover the linear suppression on small scales as expected from theoretical considerations mentioned in section \ref{sec:neutrino_physis}. Moreover, we find the anticipated additional suppression of power on small scales ($ k \sim 0.8 h/\si{Mpc}$), which is a result of the different expansion histories in both cosmologies: Due to a slower structure growth, matter perturbations will also enter the non-linear regime slightly later, which then causes an extra reduction in the non-linear boost of the spectrum \cite{Lesgourgues:2013sjj}. 

Despite the differences in the initial conditions generation and the N-body algorithms, both codes produce very similar results. Although not visible in the plot, the 2LPT initial conditions in \GADGET{} enhance the $z=0$ power spectrum on small scales by a few percent. However, as long as the same order of LPT is used for the reference cosmology as well as the massive neutrino cosmology, the difference factors out and the relative power spectrum converges to the same solution. Similarly, the sampling variance on large scales, that is clearly visible in the power spectrum (upper panel), gets factored out in the relative power spectrum. In general, as long as the same initial seed and force computation method is used, the relative power spectrum is much less sensitive to the accuracy of the underlying N-body simulation which makes it well suited to single out the accuracy of the treatment of neutrinos.

\begin{figure}[t]
\centering
  \includegraphics[width=0.98\linewidth]{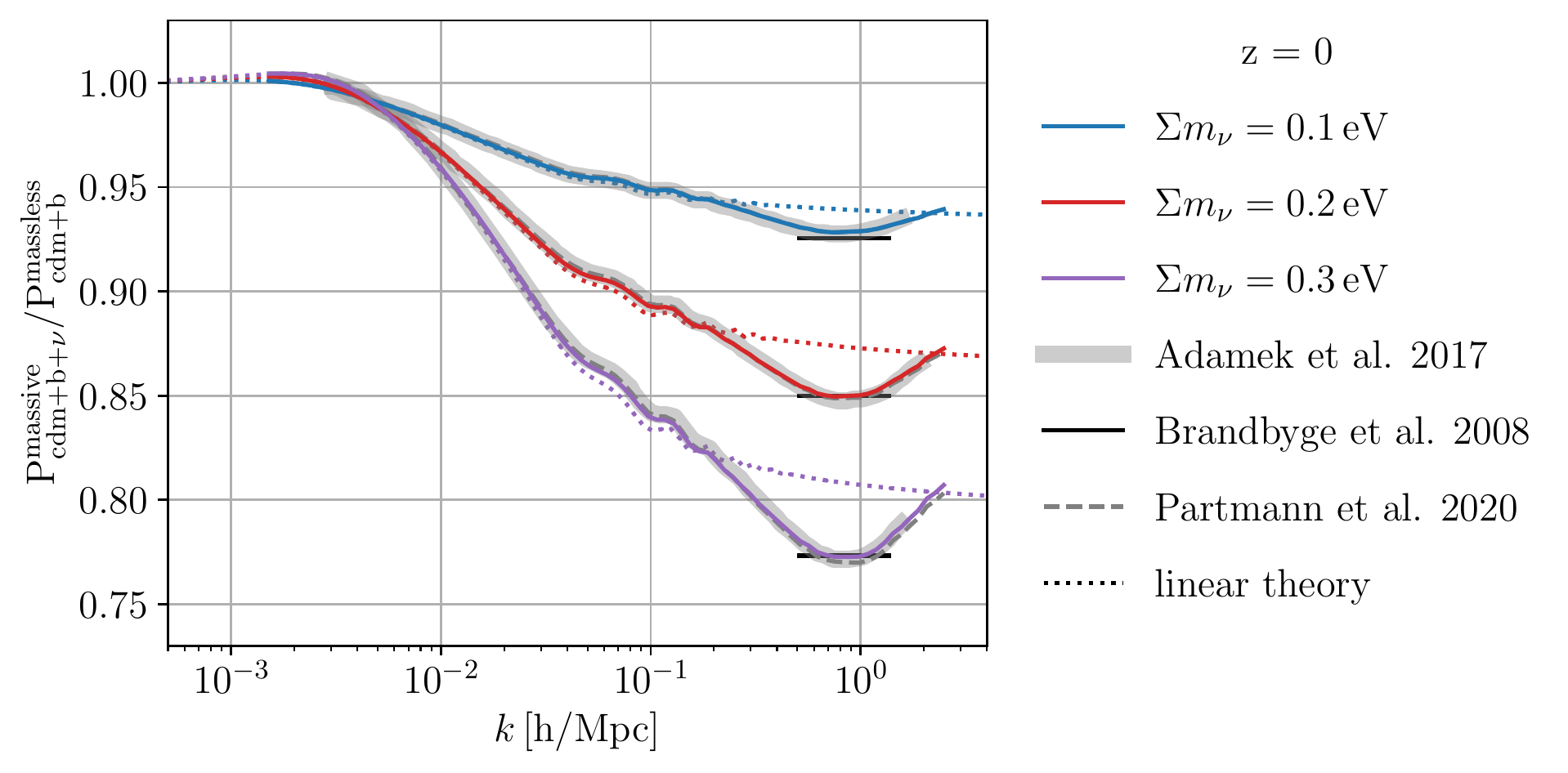}
 \caption{Comparison of cosmologies with different neutrinos masses to our reference cosmology. We also show the results of the fully relativistic and non-linear simulations in \cite{Adamek_2017} and find excellent agreement. The black lines represent the fitting function $\Delta P / P = -9.8f_{\nu}$ presented in \cite{Brandbyge_2008}. Results obtained with our previous method (dashed line) \cite{Partmann_2020} are almost indistinguishable from the improved implementation.}
 \label{fig:different_masses}
\end{figure}

To identify the range of validity of our method, we run simulations with different neutrino mass sums $\sum m_{\nu} = \{ 0.1 ,0.2, 0.3 \} \si{.eV}$ and compare the results to the well established, fully non-liner, simulations in \cite{Adamek_2017}. Since our method only generates the CDM+b power spectrum while the power spectra in \cite{Adamek_2017} also include the density contribution from neutrinos, we add a linear representation of the neutrino density field at final time. Because the neutrino power spectrum is suppressed by several orders of magnitude on small scales (e.g. 4 orders of magnitude at $k \sim 1 \si{.Mpc}^{-1}$), the assumption of linear neutrino densities does not cause a loss of accuracy. The results for the suppression of the total (CDM+b+$\nu$) power spectrum are shown in figure \ref{fig:different_masses}. The grey band shows the results from \cite{Adamek_2017}, and the black bars show the fitting function $\Delta P / P = -9.8f_{\nu}$ . The comparison to \cite{Adamek_2017} clearly shows the accuracy of our method in the tested regime. Our new method also agrees with our previous method \cite{Partmann_2020}, that used the Newtonian motion gauge method, but without the backscaling prescription.

\section{Conclusions}
\label{sec:Conclude}

In this paper we present a novel Newtonian motion gauge suitable for massive neutrino cosmologies, allowing us to merge the output of a Newtonian N-body simulation of cold dark matter with a run of an Einstein-Boltzmann solver to obtain a consistent non-linear and relativistic simulation of dark matter including the impact of massive neutrinos. Our method is not an approximate treatment of neutrinos, but exact to the given order in cosmological perturbation theory.

Compared to previous work \cite{Partmann_2020} our new approach has several important improvements. For the first time we construct a backwards Newtonian motion gauge, meaning that the Newtonian simulation can be started from the commonly employed backscaling initial conditions and does not rely on a specific tailor made initialisation. Just as in a massless neutrino case, we show that the synchronous gauge matter power spectrum of cold dark matter and baryons can be used at the present time and scaled back by the linear growth function. Note that we do not include the power spectrum of massive neutrinos in backscaling, as neutrinos are not represented by N-body particles in our approach.

We have further, for the first time, constructed a Nm gauge such that the corresponding N-body simulation contains only a pure growing mode, with all decaying modes present in the actual Universe being described via our choice of gauge. It is therefore no longer an approximation to employ pure growing mode initial conditions in N-body simulations, but theoretically justified in our framework. In contrast to our previous work this allows the use of 2LPT or higher order initial conditions, that are based on a Newtonian growing mode, out of the box. It may further be an interesting approach for studies of decaying modes independent of massive neutrinos.

Our N-body simulations are in most parts equivalent to ordinary Newtonian N-body simulations of cold dark matter and we only introduce three simple steps to make them compatible with a massive neutrino cosmology. First, we modify the N-body code to include the impact of massive neutrinos in the background evolution, i.e. in the Hubble rate. Second, we use the above described initial conditions, either with the the Zel'dovich approximation or the 2LPT formalism to realise the cold species only. We then run the simulation without further modifications, which in particular allows us to take full advantage of the highly optimised codes out-of-the-box. The simulation particles represent the evolution of cold dark matter and baryons in the Nm gauge coordinates. The final step is a post-processing to translate the output to a more well-known gauge for analysis. We find that in many cases this correction remains small, especially at late times since the present-day correction vanishes by design when using backscaling initial conditions. Most of the impact that massive neutrinos have on the dark matter evolution is already included in our simulation by the change of the Hubble rate and the use of our backscaling initial conditions. 

Using those minimal modifications, we enable Newtonian N-body codes to include the impact of linear massive neutrinos on the growth of structure. We validate our method with two different N-body codes (\gevolution{} and \GADGET) and for three different neutrinos masses $\sum m_{\nu} = 0.1 , 0.2 , 0.3 \si{.eV}$. Our results are compatible with the fully non-linear simulations carried out in \cite{Adamek_2017}, that rely on a particle based implementation of massive neutrinos.

Beyond the sampling noise of the N-body simulation, our precision is limited by neglecting higher order perturbations of the relativistic potential and neutrino density perturbations. However, treating neutrinos perturbatively does not inflict a significant loss of accuracy, since non-linearities in the neutrino sector are strongly suppressed in the allowed mass range. We further show that all metric potentials in our Nm gauge remain perturbatively small and stable at all times. The overall precision is thus approximated by the precision of N-body simulation itself.

Finally we present an updated strategy for backscaling, which is relevant for any N-body simulation employing backscaling initial conditions. To better tailor N-body simulations to a particular survey, backscaling does not need to be performed from the present-day power spectrum. Instead, one may use a power spectrum from roughly the smallest redshift at which data is being collected, see appendix \ref{altImp} for more details. In that case, the need to post-process the output to include relativistic corrections or the impact of massive neutrinos is minimised. For the currently favoured neutrino masses and surveys, a post-processing of the output can be completely avoided without sacrificing much accuracy. 

In summary, a simulation of CDM+b in a massive neutrino cosmology using our framework is in fact as simple and efficient as a simulation in the radiation-free case, with no computational overhead or loss in accuracy. Furthermore our method is already relativistic by design and includes the corresponding large-scale corrections by default.

\newpage

\appendix

\section{Computation of $\HT$}
\label{sec:compHT}
As outlined in section \ref{sec:bwNmgauge}, we can construct the metric potential $\HT$ as a solution of a second order ODE \eqref{eqn:H_T_ode} using the semi-analytical solution \eqref{eqn:H_T_ansatz} and \eqref{eqn:C_plus_minus}. In practice however, it is highly non-trivial to retrieve $\HT$ in this way for practical applications. A much simpler way is to use the geometrical relation between the Newtonian and the relativistic density in the Nm gauge 
\begin{equation}
\label{eqn:geometrical_relation_app}
\delta^{\N}(\tau) = \delta^{\Nm}_{\rm m}(\tau) - \HT(\tau) + 3 \Phi(\tau) .
\end{equation}
First we note that this relation holds for every time $\tau$, and we further use that the overdensity in Nm gauge coincides with the Poisson overdensity. This equation is an algebraic equation for $\HT$ that can be used to compute $\HT$ for any Nm gauge, given that we know the value of $\delta^{\N}$. In our specific case, we want to use the backwards Nm gauge as defined in section \ref{sec:bwNmgauge}. To retrieve this solution from \eqref{eqn:geometrical_relation_app}, we need to identify $\delta^{\N}$ as the linear Newtonian overdensity obtained from backscaling,
so
\begin{equation}
\delta^{\N}(\tau) = D_+(\tau)\delta^{\N}(\tau_f),
\end{equation}
where $\delta^{\N}(\tau_f)$ is a constant. To fix this constant, we recall that we want $\HT = 3\zeta$ at final time, and we use \eqref{eqn:geometrical_relation_app} evaluated at final time. Hence we have
\begin{equation}
\HT(\tau)  = \delta^{\rm P}_{\rm m}(\tau) + 3 \Phi(\tau) - D_+(\tau) \left[  \delta^{\rm P}_{\rm m}(\tau_f) - 3\zeta(\tau_f) + 3 \Phi(\tau_f)  \right] .
\end{equation}
The big advantage of this equation is that all quantities needed to compute $\HT$ are known from Poisson gauge perturbations only. This means that this equation can quickly be implemented in a Boltzmann solver like \CLASS{}.  

\section{Alternative Implementation}
\label{altImp}
In practice we always seek to compute observables that we can compare to data. In the context of this work, we would rely on future power spectra captured by surveys like Euclid. Those surveys however are sensitive over a range of redshifts and most information is extracted from redshifts larger than zero, whereas our method is only exact without post-processing at $z=0$ by construction. But this is not a fixed property of our approach, but rather a choice that we have made by choosing to backscale from present-day power spectra. If instead we start to backscale from a higher redshift our method would be exact at that redshift, while times before or after would require corrections.

\begin{figure}[t]
\centering
  \includegraphics[width=0.9\linewidth]{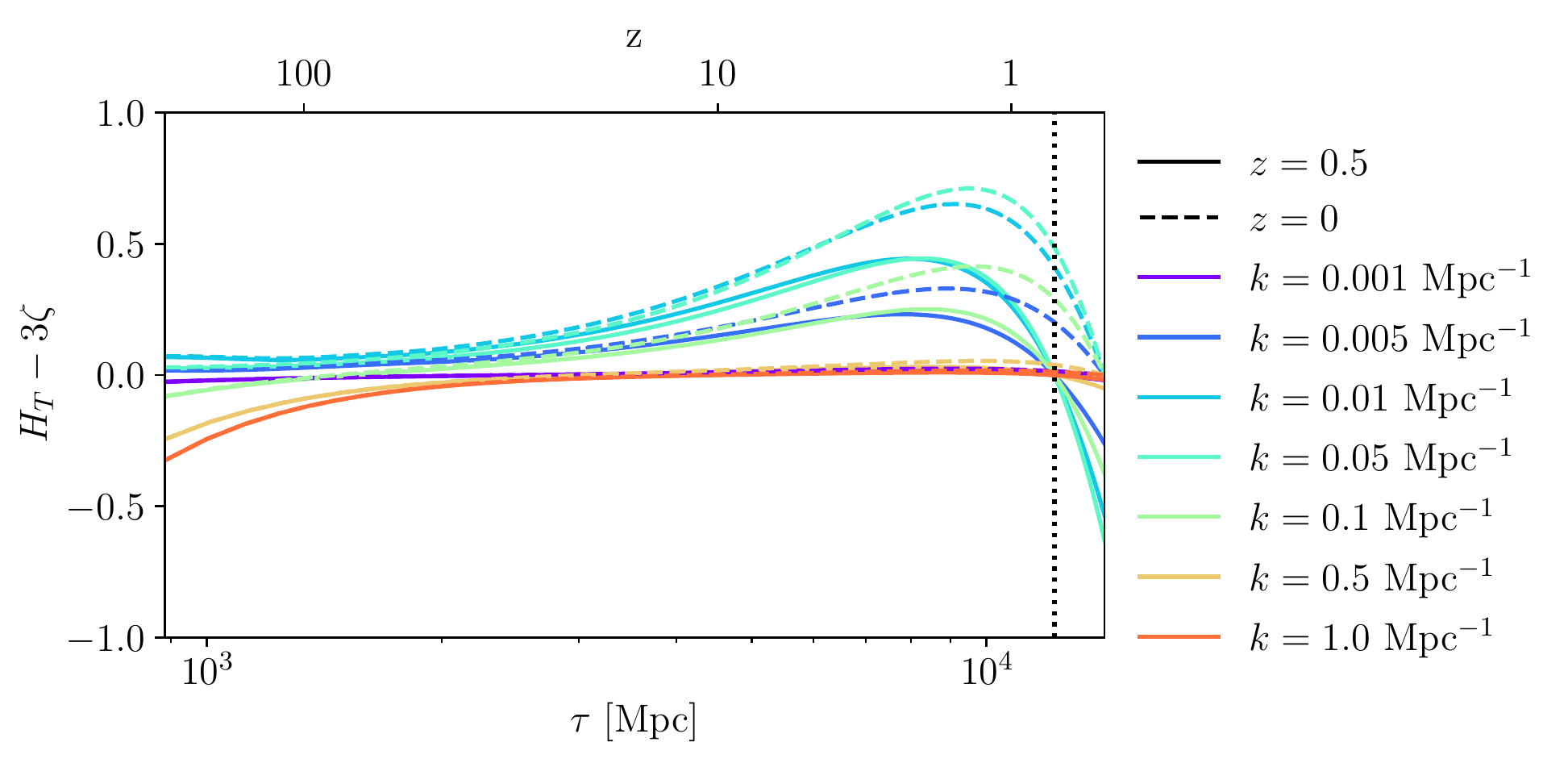}
 \caption{Shown in this figure is the potential $\HT-3\zeta$ in a $\sum m_{\nu}=0.1 \si{eV}$ cosmology. Unlike the solution presented in \ref{sec:bwNmgauge}, this solution (full lines) has the
boundary condition $\HT = 3\zeta$ at $z = 0.5$. This solution enables the resulting matter
power spectrum of the N-body simulation to be correct at $z = 0.5$. Also shown (dashed lines) is the previously presented solution which fixes $\HT = 3\zeta$ at $z = 0$.}
  \label{fig:H_T_z_05}
\end{figure}

We have checked this idea for a redshift of $z=0.5$. In figure \ref{fig:H_T_z_05} we show the metric potential $\HT$
for the Nm gauge fixed at $z = 0.5$ compared to the usual case where the present-day power spectra are employed. The overall shape of $\HT$ in those two slightly different gauges is very similar. In our novel case, the metric deviation from the N-boisson gauge remains generally smaller, especially for times $z > 0.5$. As mentioned earlier we can also approximate the error in the power spectrum, if we neglect this shift in \ref{fig:different_z_different_masses}. Form the plot it is clear that our new choice would be superior for surveys collecting most of their data before or around $z = 0.5$.

We therefore conclude that for a given survey that collects most of its data at $z > z_{\rm survey}$, backscaling initial conditions should employ the matter power spectrum at $z_{\rm survey}$ instead of the present-day spectrum. In that case the need to do a post-processing is reduced at no additional computational cost.

\bibliography{neutrino_bs}{}
\bibliographystyle{JHEP}

\end{document}